\documentclass[draftcls,12pt,onecolumn]{IEEEtran}
\pdfoutput=1 

\usepackage{amsmath,amssymb}
\usepackage{subfigure}
\usepackage{graphicx}
\usepackage{booktabs}
\usepackage{times}

\newcommand{\Xcal}{\mathcal{X}}

\newcommand{\sbi}{\mu_1(\mathcal{X}_{b}^i)}
\newcommand{\sbiSqrd}{\mu_1^2(\mathcal{X}_{b}^i)}

\newcommand{\zk}{y_k}

\newcommand{\Zz}{Y}

\newcommand{\sk}{x_k}

\newcommand{\Ss}{X}
\newcommand{\sss}{x}
\newcommand{\hk}{h_k}
\newcommand{\wk}{z_k}

\newcommand{\Ww}{Z}

\newcommand{\pmf}{P}
\newcommand{\meanConst}{\mu_1(\Xcal)}
\newcommand{\meanConstSqrd}{\mu_1^2(\Xcal)}
\newcommand{\muTwoConst}{\mu_2(\Xcal)}

\newcommand{\muTwoBisConst}{\mu_{2}'(\Xcal)}
\newcommand{\sAv}{\hat{\sss}}

\newcommand{\sAvr}{\hat{\sss}_r}
\newcommand{\sAvi}{\hat{\sss}_i}
\DeclareMathOperator{\Tr}{Tr}

\newcommand{\cov}{\text{cov}}

\newcommand{\Xcali}{\mathcal{X}_b^i}

\newcommand{\pzx}{p_{Y|X,H}(y|x,h)}
\newcommand{\pzxp}{p_{Y|X,H}(y|x',h)}

\newcommand{\capc}{{\rm C}}
\newcommand{\capCM}{{\rm C}_{\Xcal}}
\newcommand{\capCMi}{{\rm C}_{\mathcal{X}_b^i}}
\newcommand{\capBICM}{{\rm C}_{\Xcal,\gamma}}
\newcommand{\snr}{{\rm{SNR}}}
\newcommand{\units}[2]{#1\textrm{\thinspace \rm #2}}

\newcommand{\ebnoMin}{\frac{E_{\rm b}}{N_0}_{\rm min}}
\newcommand{\ebno}{\frac{E_{\rm b}}{N_0}}
\newcommand{\ebnoZ}{\frac{E_{\rm b}}{N_0}_{\rm lim}}

\newcommand{\eqdef}{\stackrel{\Delta}{=}}

\newcommand{\gf}{\nu}
\newcommand{\ca}{c_1}
\newcommand{\cb}{c_2}
\newcommand{\slope}{\zeta_0}
\newcommand{\cao}{c_{11}}
\newcommand{\cat}{c_{12}}
\newcommand{\cbo}{c_{21}}
\newcommand{\cbt}{c_{22}}

\DeclareMathOperator{\E}{E}
\DeclareMathOperator{\Ord}{O}
\DeclareMathOperator{\ord}{o}

\newcommand{\pw}{p}

\newtheorem{theorem}{Theorem}
\newtheorem{proposition}{Proposition}

\newtheorem{corollary}{Corollary}

\graphicspath{{figures/}}

\begin{document}

\title{Bit-Interleaved Coded Modulation in the Wideband Regime}


\author{Alfonso Martinez, Albert Guill{\'e}n i F{\`a}bregas, Giuseppe Caire and Frans Willems
   \thanks{A. Martinez and F. Willems are with the Department of Electrical Engineering, Technische Universiteit Eindhoven, Postbus 513, 5600 MB Eindhoven, The Netherlands, e-mail: \tt alfonso.martinez@ieee.org, f.m.j.willems@tue.nl.}
    \thanks{A. Guill{\'e}n i F{\`a}bregas is with the Department of Engineering, University of Cambridge, Cambridge, CB2 1PZ, UK, e-mail: \tt guillen@ieee.org.}
    \thanks{G. Caire is with the Electrical Engineering Department, University of Southern California, 3740 McClintock Ave.,\, Los Angeles, CA 90080, USA, e-mail: \tt caire@usc.edu.}
  \thanks{This work has been presented in part at the 2007 International Symposium on Information Theory and Applications, ISIT 2007, Nice (France), June 2007.}
  \thanks{This work has been partly supported by the International Incoming Short Visits Scheme 2007/R2 of the Royal Society and by the Australian Research Council under ARC grant
DP0558861.}
}

\maketitle

\begin{abstract}
The wideband regime of bit-interleaved coded modulation (BICM) in Gaussian channels is studied.
The Taylor expansion of the coded modulation capacity for generic signal constellations at low signal-to-noise ratio (SNR) is derived and used to determine the corresponding expansion for the BICM capacity.
Simple formulas for the minimum energy per bit and the wideband slope are given.
BICM is found to be suboptimal in the sense that its minimum energy per bit can be larger than the corresponding value for coded modulation schemes. The minimum energy per bit using standard Gray mapping on $M$-PAM or $M^2$-QAM is given by a simple formula and shown to approach \units{-0.34}{dB} as $M$ increases. Using the low SNR expansion, a general trade-off between power and bandwidth in the wideband regime is used to show how a power loss can be traded off against a bandwidth gain.
\end{abstract}

\newpage
\section{Introduction and Motivation}
\begin{figure}[htbp]
  \centering
  \subfigure[\label{fig:c_bicm_gr_sp_snr} Capacity as a function of $\snr$.]{\includegraphics[width=0.7\columnwidth]{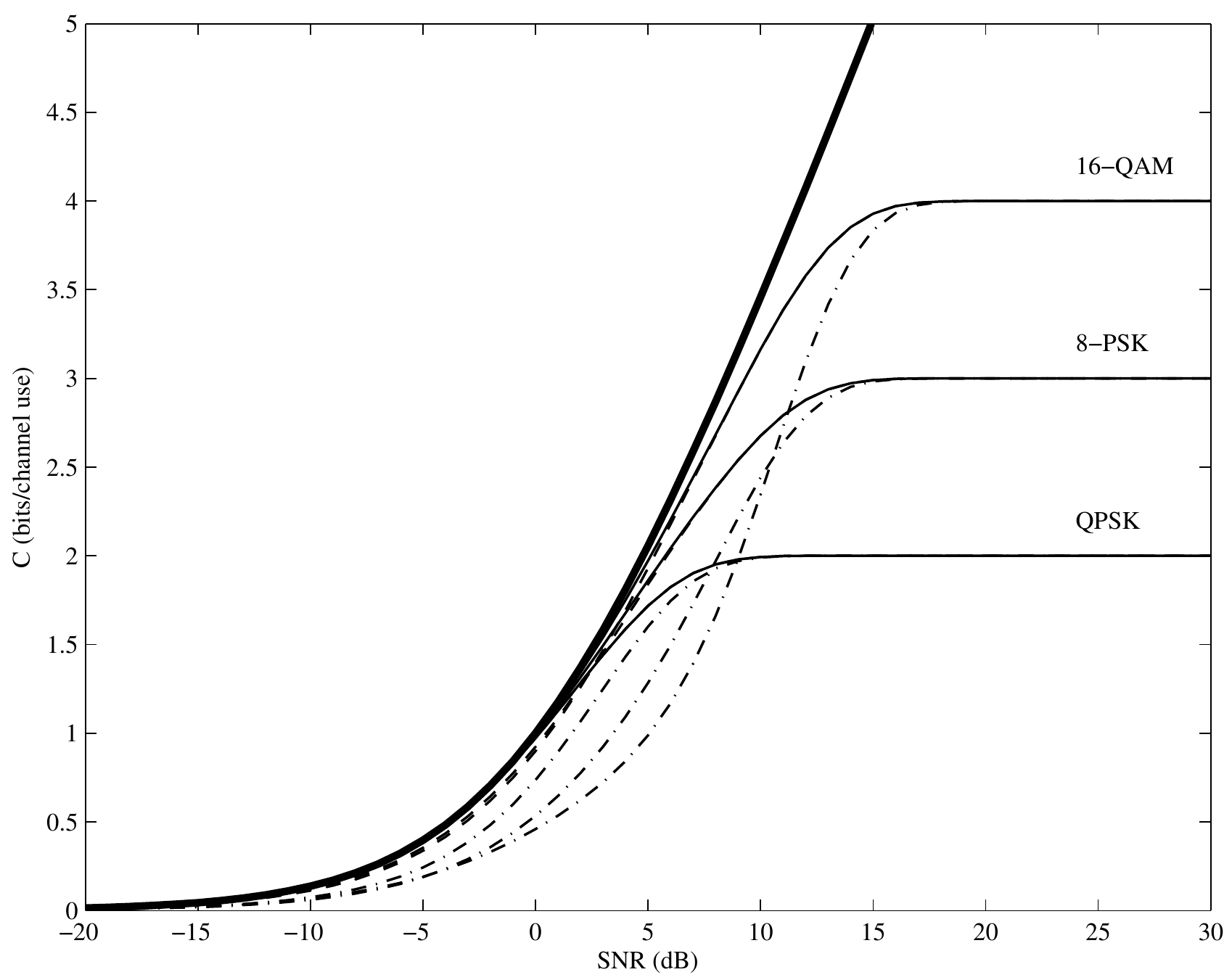}}
   \subfigure[\label{fig:c_bicm_gr_sp_ebno} Capacity as a function of $\ebno$.]{ \includegraphics[width=0.7\columnwidth]{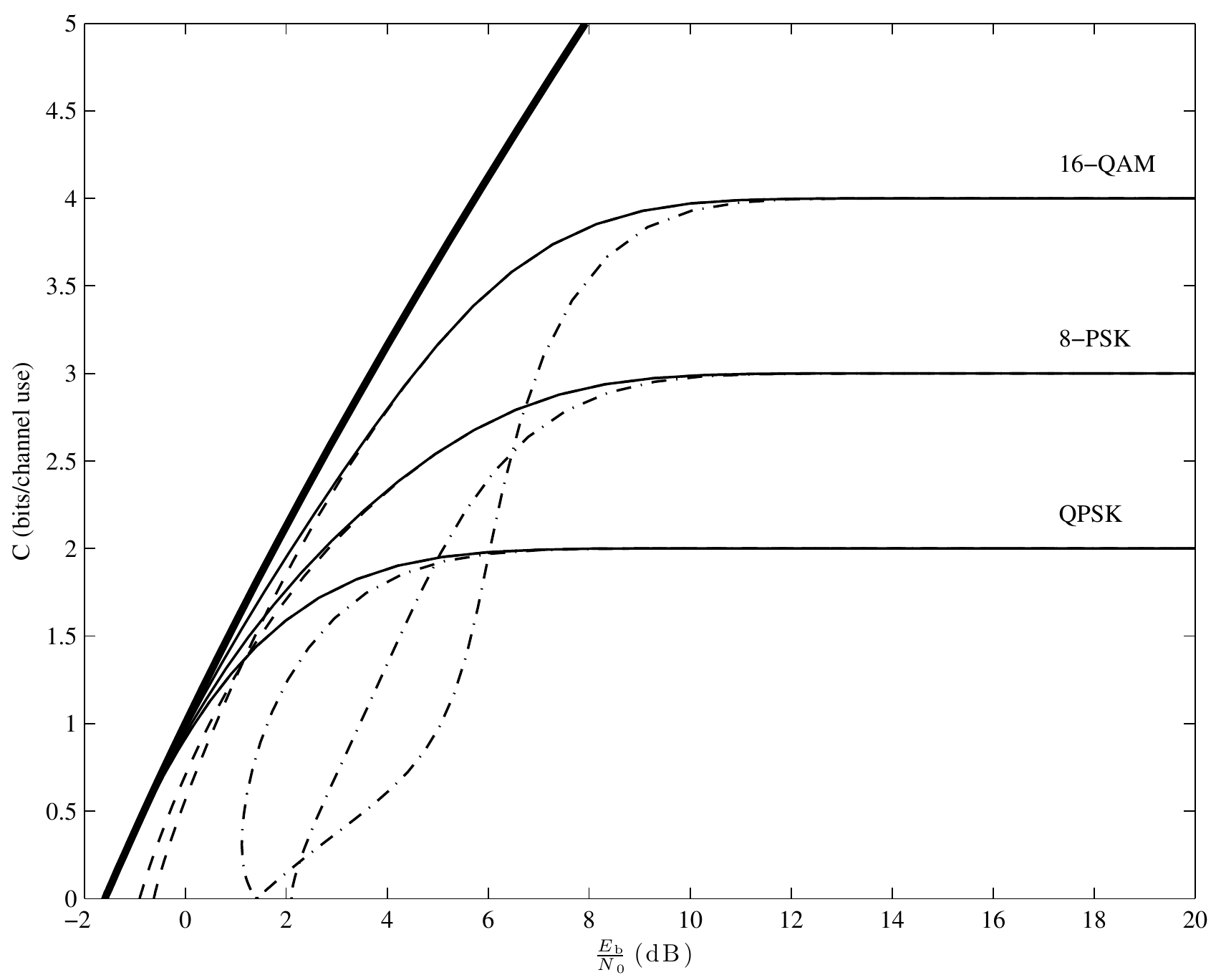}}
  \caption{Channel capacity (in bits per channel use) with memoryless binary labeling and BICM-ML decoding for multiple signal constellations with uniform inputs in the AWGN channel. Gray and set partitioning labeling rules correspond to thin dotted and dashed-dotted lines respectively. For reference, the capacity with Gaussian inputs is shown in thick solid lines and the CM channel capacity with uniform inputs \eqref{eq:capCMAWGN} with thin solid lines.}
  \label{fig:c_bicm_awgn}
\end{figure}

Bit-interleaved coded modulation (BICM) was originally proposed by Zehavi \cite{zehavi1992ptc} and further elaborated by Caire {\em et al.} \cite{caire1998bic} as a practical way of constructing efficient coded modulation schemes over non-binary signal constellations.
Reference \cite{caire1998bic} defined and computed the channel capacity of BICM under a sub-optimal {\em non-iterative} decoder, and compared it to the coded modulation capacity, assuming equiprobable signalling over the constellation.
When natural reflected Gray mapping was used, the BICM capacity was found to be near optimal at high signal-to-noise ratio (see Figure \ref{fig:c_bicm_gr_sp_snr}). Nevertheless, plots of the BICM capacity as a function of the energy per bit for reliable communication (see Figure \ref{fig:c_bicm_gr_sp_ebno}) reveal the suboptimality of BICM with the non-iterative decoder of \cite{zehavi1992ptc,caire1998bic} for low rates, that is in the power-limited or {\em wideband} regime.

Recent work by Verd\'u \cite{verdu2002sew} presents a detailed treatment of the wideband regime. He studied the minimum bit energy-to-noise ratio $\ebnoMin$ for reliable communication and the wideband slope, i.e., the first-order expansion of the capacity for low $\ebnoMin$, under a variety of channel models and channel state information (CSI) assumptions. These results are obtained by using a second-order expansion of the channel capacity at zero signal-to-noise ratio (SNR). Furthermore, using these results, he obtained a general tradeoff between data rate, power and bandwidth in the wideband regime. In particular, Verd\'u\cite{verdu2002sew} studied the bandwidth penalty incurred by using suboptimal signal constellations in the low-power regime. An implicit assumption of this tradeoff was that the power cannot change together with the bandwidth.

Motivated by the results of Figure \ref{fig:c_bicm_gr_sp_ebno} and by Verd\'u's analysis \cite{verdu2002sew}, in this paper, we give an analytical characterization of the behaviour of BICM in the low-power regime. Studying the behaviour of BICM at low rates may prove useful in the design of multi-rate communication systems where rate adaptation is carried out by modifying the binary code, while keeping the modulation unchanged.
In the process, we derive a number of results of independent interest for coded modulation over the Gaussian channel. In particular, the first two coefficients of the Taylor expansion of the coded modulation capacity for arbitrary signal constellations at zero SNR are derived, and used to obtain the corresponding coefficients for BICM. We also obtain a closed form expression for the minimum $\ebno$ for BICM using QAM constellations with natural reflected Gray mapping, and we show that for large constellations it approaches \units{-0.34}{dB}, resulting in a \units{1.25}{dB} power loss with respect to coded modulation. Using these results, we derive  the trade-off between power and bandwidth in the wideband regime that generalizes the results of \cite{verdu2002sew} to capture the effects of changing both power and bandwidth.

This paper is organized as follows. Section \ref{sec:model} introduces the system model, basic assumptions and notation. Section \ref{sec:wideband} defines the wideband regime, and presents the low-SNR expansion for both coded modulation and BICM. Section \ref{sec:bandwidthPower} introduces the general trade-off between power and bandwidth. Concluding remarks appear in Section \ref{sec:conclusions}. Proofs of various results are in the Appendices.

\section{Model and Assumptions}
\label{sec:model}
We consider a complex-valued, discrete-time additive Gaussian noise channel with fading. The $k$-th channel output $\zk$ is given by
\begin{equation}\label{eq:channelAWGN}
  \zk = \hk\,\sqrt{\snr} \,\sk+ \wk,
\end{equation}
where $\sk$ is the $k$-th channel input, $\hk$ a fading coefficient, and $\wk$ an independent sample of circularly symmetric complex-valued Gaussian noise of unit variance; $\snr$ denotes the average signal-to-noise ratio at the receiver. The transmitted, received, noise and fading samples, are realizations of the random variables $X,Y,Z$ and $H$.
The fading coefficients $\hk$ are independently drawn from a density $\pw_H(\hk)$ and are assumed known at the receiver. For future use we define the squared magnitudes of the fading coefficients by $\chi_k=|h_k|^2$.
For a given fading realization $\hk$, the conditional output probability density is given by
\begin{equation}\label{eq:pChannelOutput}
    \pw_{\Zz|\Ss,H}(\zk|\sk,\hk) = \frac{1}{\pi}e^{-|\zk-\hk\sqrt{\snr}\,\sk|^2}.
\end{equation}

The channel inputs are modulation symbols drawn from a constellation set $\Xcal$ with probabilities $\pmf_{\Ss}(\sss)$. We denote the cardinality of the constellation set by $M = |\Xcal|$ and by $m = \log_2 M$ the number of bits required to index a modulation symbol. We define the constrained capacity $\capCM$ (or coded modulation capacity) as the corresponding mutual information between channel input and output, namely
\begin{align}\label{eq:capCMAWGN}
    \capCM(\snr) &= -\E \Biggl[\log \Biggl(\sum_{\sss'\in\Xcal}{\pmf_{\Ss}(\sss')e^{-|H\sqrt{\snr}(\Ss-\sss')+Z|^2+|Z|^2}}\Biggr)\Biggr]
\end{align}
where the expectation is performed over $X, Z$ and $H$. If the symbols are used with equal probabilities, i.\ e.\ $\pmf_{\Ss}(\sss) = M^{-1}$, we refer to the constrained capacity as uniform capacity, and denote it by $\capCM^\text{u}$.

As we will see later, it proves convenient to consider general constellation sets with arbitrary first and second moments, respectively denoted by $\meanConst$ and $\muTwoConst$, and given by
\begin{align}
    \meanConst &\triangleq \E[X] = \sum_{\sss\in\Xcal} \sss\pmf_{\Ss}(\sss), \notag\\
    \muTwoConst &\triangleq \E[|X|^2]= \sum_{\sss\in\Xcal} |\sss|^2\pmf_{\Ss}(\sss).\notag
\end{align}
Practical constellations have zero mean, i.\ e.\ $\meanConst = 0$, and unit energy, that is $\muTwoConst = 1$.

In order to transmit at rates close to the coded modulation capacity, multi-level coding or non-binary codes are needed \cite{forneyjr1998mac,wachsmann1999mct}.
Alternatively, in bit-interleaved coded modulation (BICM) binary codes are mapped with a binary mapping rule $\gamma$ onto non-binary modulations \cite{zehavi1992ptc,caire1998bic}. Caire {\em et al.} found that BICM with natural reflected Gray mapping and low-complexity non-iterative demodulation attains very good performance, close to that of coded modulation with equiprobable signalling \cite{caire1998bic}.
For infinite interleaving, the channel is separated into a set of $m$ parallel independent subchannels, and one defines the so-called BICM capacity, denoted by $\capBICM$, given by
\begin{align}
    \capBICM(\snr) &= \sum_{i=1}^m I(B_i;\Zz)\\
    &= \sum_{i=1}^m \E \Biggl[\log \frac{{\sum_{\sss'\in\Xcal_b^i} e^{-|H\sqrt{\snr}(\Ss-\sss')+\Ww|^2}}}{\frac{1}{2}\sum_{\sss'\in\Xcal} e^{-|H\sqrt{\snr}(\Ss-\sss')+\Ww|^2}}\Biggr]
\end{align}
where $B_i$ denotes the binary input random variable corresponding to the $i$-th parallel channel (see \cite{caire1998bic} for details), $\Xcali$ are the sets of constellation symbols with bit $b$ in the $i$-th position of the binary label and the expectation is performed over all input symbols $\sss$ in $\Xcali$ for $b= 0,1$, and over all possible noise and fading realizations, respectively $\Ww$ and $H$.
An equivalent, yet alternative, definition is given by the following.
\begin{proposition}\label{prop:proofCBICM}
The BICM capacity can be expressed as
\begin{align}\label{eq:capBICM}
    \capBICM &= \sum_{i=1}^m \frac{1}{2}\sum_{b=0,1}(\capCM^\text{u}-\capCMi^\text{u}),
\end{align}
where $\capCM^\text{u}$ and $\capCMi^\text{u}$ are, respectively, the constrained capacities for equiprobable signalling in $\Xcal$ and $\Xcali$.
\end{proposition}
\begin{proof}
The proof is given in Appendix \ref{app:proofCBICM} \footnote{This expression has been independently derived in \cite{bran2007clas}.}.
\end{proof}
In general, the sets $\Xcali$ have non-zero mean and non-unit average energy.
This result reduces the analysis of the BICM capacity to that of coded modulation over constellation sets with arbitrary first and second moments.

\section{Wideband Regime}
\label{sec:wideband}
In the wideband regime, as defined by Verd{\'u} in \cite{verdu2002sew}, the energy of a single bit is spread over many channel degrees of freedom, resulting in a low signal-to-noise ratio $\snr$. It is then convenient to study the asymptotic behavior of the channel capacity as $\snr\to 0$. In general, the capacity\footnote{This capacity may be the coded modulation capacity, or the BICM capacity.} (in nats per channel use) admits an expansion in terms of $\snr$,
\begin{align}
    \capc(\snr) = \ca\snr +\cb\snr^2 + \ord\bigl(\snr^2\bigr),
    \label{eq:c_snr}
\end{align}
where $\ca$ and $\cb$ depend on the modulation format, the receiver design, and the fading distribution.

Among the several uses for the coefficients $\ca$ and $\cb$, Verd{\'u} \cite{verdu2002sew} studied the transformation of expansion~\eqref{eq:c_snr} into a function of the bit-energy to noise ratio $\ebno$,
\begin{equation}
    \ebno = \frac{\snr}{\capc \log_2 e}.
    \label{eq:snr_eta_ebno}
\end{equation}
In linear scale for $\ebno$, one obtains
\begin{equation}
\capc\left(\ebno\right) = \slope \left(\ebno - \ebnoZ\right) + \Ord\left(\left(\Delta\ebno\right)^2\right)
\label{eq:c_ebno}
\end{equation}
where $\Delta\ebno\eqdef \ebno-\ebnoZ$ and
\begin{align}
\slope \eqdef -\frac{\ca^3}{\cb\log^2 2}, ~~~~~\ebnoZ \eqdef \frac{\log2}{\ca}.
\end{align}
The parameter $\slope$ is Verd{\'u}'s wideband slope in linear scale \cite{verdu2002sew}. We avoid using the word minimum for $\ebnoZ$, since there exist communication schemes with a negative slope $\slope$, for which the absolute minimum value of $\ebno$ is achieved at non-zero rates. In these cases, the expansion at low power is still given by Eq. \eqref{eq:c_ebno}.
The derivation of Eq.~\eqref{eq:c_ebno} can be found in Appendix~\ref{app:proofAsymptotic}.

A second important use of the coefficients $\ca$ and $\cb$ was the analysis of the bandwidth penalty incurred by using suboptimal constellations in the low-power regime \cite{verdu2002sew}. An implicit assumption in \cite{verdu2002sew} was the power cannot change together with the bandwidth. In Section~\ref{sec:bandwidthPower} we relax this assumption and give a formula for the trade-off between power penalty and bandwidth penalty and apply it to compare BICM with standard coded modulation.

In the following, we determine the coefficients $\ca$ and $\cb$ in the expansion~\eqref{eq:c_snr} for generic constellations, and use them to derive the corresponding results for BICM. Before proceeding along this line, we note that Theorem~12 of \cite{verdu2002sew} covers the effect of fading.
The coefficients $\ca$ and $\cb$ for a general fading distribution are
\begin{align}
    \ca = \E[\chi]\ca^{\text{AWGN}}, \qquad
    \cb = \E[\chi^2]\cb^{\text{AWGN}},
\end{align}
where the coefficients $\ca^{\text{AWGN}}$ and $\cb^{\text{AWGN}}$ are in absence of fading. Hence, even though we focus only on the AWGN channel, all results are valid for general fading distributions.

\subsection{Coded Modulation}

For the unconstrained case, where the capacity is $\log(1+\snr)$, then $\ca = 1$ and $\cb = -\frac{1}{2}$.
In \cite{prelov04:soa}, Prelov and Verd{\'u} determined the coefficients $\ca$ and $\cb$ for the so-called proper-complex constellations introduced by Neeser and Massey \cite{neeser93pcr}, which satisfy
\[
\muTwoBisConst \triangleq \E[X^2] = \sum_{\sss\in\Xcal} \sss^2\pmf_{\Ss}(\sss) = 0,
\]
where $\muTwoBisConst$ is a second-order pseudo-moment, borrowing notation from the paper \cite{neeser93pcr}.
The coefficients for coded modulation formats with arbitrary first and second moments are given by the following result.
\begin{theorem}\label{prop:c1c2CM}
Consider coded modulation schemes over a signal set $\Xcal$ used with probabilities $\pmf_{\Ss}(\sss)$ in the Gaussian channel. Then,
the first two coefficients of the Taylor expansion of the constrained capacity $\capCM(\snr)$ around $\snr = 0$ are given by
\begin{align}\label{eq:c1c2CM-1}
\ca &= \muTwoConst - \bigl|\meanConst\bigr|^2 \\
\cb &= -\frac{1}{2} \Bigl(\bigl(\muTwoConst - \bigl|\meanConst|^2\bigr)^2+\bigl|\muTwoBisConst-\meanConstSqrd\bigr|^2\Bigr).\label{eq:c1c2CM-2}
\end{align}

When $\meanConst = 0$ (zero mean) and $\muTwoConst = 1$ (unit energy),
\begin{align}
\ca = 1,\qquad
\cb = -\frac{1}{2} \Bigl(1+\bigl|\muTwoBisConst\bigr|^2\Bigr),
\end{align}
and the bit-energy-to-noise ratio at zero $\snr$ is $\ebnoZ = \log 2$.
\end{theorem}
\begin{proof}
See Appendix \ref{app:cmGaussian-prelov}.
\end{proof}
The formula for $\ca$ is known, and can be found as Theorem~4 of \cite{verdu2002sew}. Also, for proper-complex constellations $\cb = -\frac{1}{2}$, as found in \cite{prelov04:soa}.
The second-order coefficient is bounded by $-1 \leq \cb \leq -\tfrac{1}{2}$,
the maximum ($\cb = -1/2$) being attained when the constellation has uncorrelated real and imaginary parts and the energy is equally distributed among the real and imaginary parts.

Applied to some practical signal constellations with equiprobable symbols, Theorem~\ref{prop:c1c2CM} gives the following corollaries, whose respective proofs are straightforward.

\begin{corollary}\label{cor:pskBeta}
For uniform $M$-PSK, $\cb = -1$ if $M = 2$ and $\cb = -\frac{1}{2}$ if $M > 2$.
\end{corollary}

This result extends Theorem~11.1 of \cite{verdu2002sew}, where the result held for QPSK, a simple example of proper-complex constellation.
\begin{corollary}
When $\Xcal$ represents a mixture of $N$ uniform $M_n-$PSK constellations for $n=1,\dotsc,N$, $\cb^{} = -\frac{1}{2}$ if and only if $M_n>2$ for all rings/sub-constellations $n=1,\dotsc,N$.
\end{corollary}
This applies to APSK modulations, for instance. In \cite{verdu2002sew} Theorem~11.2 stated the result for mixtures of QPSK constellations.

\subsection{Bit-Interleaved Coded Modulation}

First, for fixed label index, $i$, and bit value $b$, let us respectively define the quantities $\sbi$, $\mu_2(\Xcali)$, and $\mu_{2}'(\Xcali)$, as the mean, the second moment, and the average of the squared symbols in the set $\Xcali$.
Then, we have the following.
\begin{theorem}\label{prop:c1c2BICM}
Assume a constellation set $\Xcal$ with zero mean and unit average energy. The coefficients $\ca$ and $\cb$ for the BICM capacity $\capBICM$
are given by
\begin{align}
    \ca &= \sum_{i=1}^m\frac{1}{2}\sum_b |\sbi|^2,\\
    \cb &= \sum_{i=1}^m \frac{1}{4} \sum_{b=0,1} \biggl(\Bigl(\mu_2(\Xcali)-|\sbi|^2\Bigr)^2-\bigl(1+|\muTwoBisConst|^2\bigr)+\bigl|\mu_{2}'(\Xcali)-\sbiSqrd\bigr|^2\biggr).
\end{align}
\end{theorem}

\begin{proof}
See Appendix \ref{app:c1c2BICM}.
\end{proof}

Table~\ref{tab:ebnoMin-CM} reports the numerical values for the coefficients $\ca$ and $\cb$, as well as the bit signal-to-noise ratio $\ebnoZ$ and wideband slope $\slope$ for various cases, namely QPSK, 8-PSK and 16-QAM modulations and Gray and Set Partitioning (anti-Gray for QPSK) mappings.
\begin{table}[htbp]
  \centering
  \caption{$\ebnoZ$ and Wideband Slope coefficients $\ca,\cb$ for BICM in AWGN.}
  \begin{tabular}{@{}lrrrrrrrrrrrr@{}}
    \toprule
    & \multicolumn{6}{c}{Modulation and Mapping} \\ \cmidrule(l){2-7}
   & \multicolumn{2}{c}{QPSK}  & \multicolumn{2}{c}{8-PSK}  & \multicolumn{2}{c}{16-QAM} \\ \cmidrule(l){2-7}
     & GR & A-GR & GR & SP & GR & SP \\ \cmidrule(l){2-7}
    $\ca$ & 1.000 & 0.500 & 0.854 & 0.427 & 0.800 & 0.500 \\
    $\ebnoZ$ & 0.693 & 1.386 & 0.812 & 1.624 & 0.866 & 1.386 \\
    $\ebnoZ$ (dB) & -1.592 & 1.419 & -0.904 & 2.106 & -0.627 & 1.419 \\
    \cmidrule{1-7}
    $\cb$ & -0.500 & 0.250 & -0.239 & 0.005 & -0.160 & -0.310 \\
    $\slope$ & 4.163 & -1.041 & 5.410 & -29.966 & 6.660 & 0.839 \\
    \bottomrule
  \end{tabular}
  \label{tab:ebnoMin-CM}
\end{table}

\begin{figure}[htbp]
  \centering
  \includegraphics[width=0.9\columnwidth]{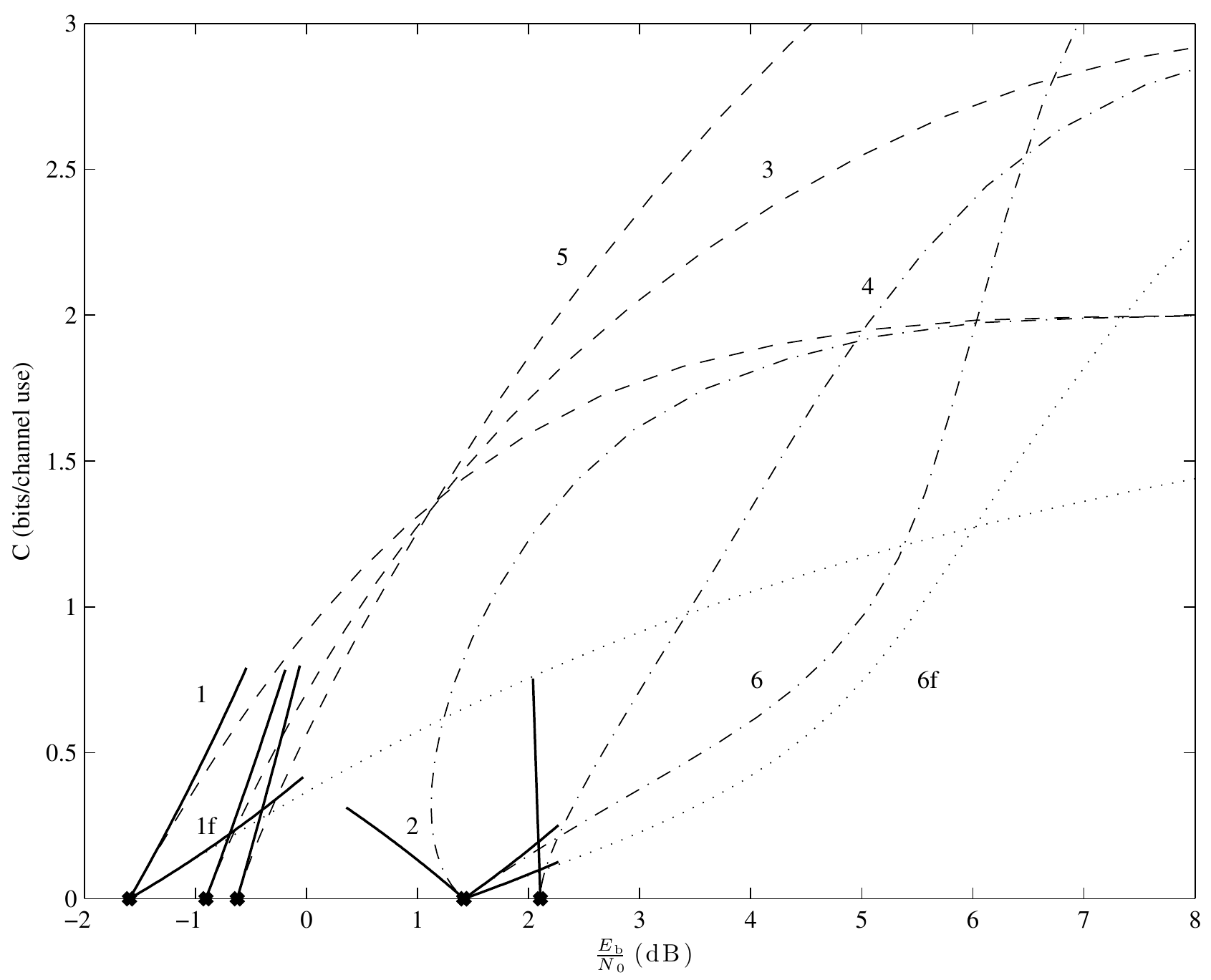}
  \caption{BICM capacity (in bits per channel use). Labels 1 and 2 are QPSK, 3 and 4 are 8-PSK and 5 and 6 are 16-QAM. Gray and set partitioning labeling rules correspond to dashed (and odd labels) and dashed-dotted lines (and even labels) respectively. Dotted lines are cases 1 and 6 with Nakagami-$0.3$ and Nakagami-$1$ (Rayleigh) fading (an `f' is appended to the label index). Solid lines are linear approximation around $\ebnoZ$.}
  \label{fig:c_bicm_awgn_zoom}
\end{figure}

In Figure~\ref{fig:c_bicm_awgn_zoom}, the approximation in Eq.~\eqref{eq:c_ebno} is compared with the capacity curves. As expected, a good match for low rates is observed. We use labels to identify the specific cases: labels 1 and 2 are QPSK, 3 and 4 are 8-PSK and 5 and 6 are 16-QAM. Also depicted is the linear approximation to the capacity around $\ebnoZ$, given by Eq.~\eqref{eq:c_ebno}. Two cases with Nakagami fading are also included in Figure~\ref{fig:c_bicm_awgn_zoom}, which also show good match with the estimate, taking into account that $\E[\chi] = 1$ and $\E[\chi^2] = 1+1/\gf$ for Nakagami-$\nu$ fading. An exception is 8-PSK with set-partitioning, for which the approximation is valid for a very small range of rates, since $\cb$ is positive and very small, which implies a very large slope.

In general, it seems difficult to draw general conclusions for arbitrary mappings from Theorem~\ref{prop:c1c2BICM}. A notable exception, however, is the analysis under natural reflected Gray mapping.
\begin{theorem}\label{thm:alphaBICMGray}
    For $M$-PAM and $M^2$-QAM and natural, binary-reflected Gray mapping,
    the coefficient $\ca$ in the Taylor expansion of the BICM capacity $\capBICM$ at low $\snr$ is
    \begin{equation}
        \ca = \frac{3\cdot M^2}{4(M^2-1)},
    \end{equation}
    and the minimum $\ebnoZ$ is
    \begin{equation}\label{eq:alphaBICMGray}
        \ebnoZ = \frac{4(M^2-1)}{3\cdot M^2}\log 2.
    \end{equation}

    As $M\to\infty$, $\ebnoZ$ approaches
      $\frac{4}{3}\log 2 \simeq \units{-0.3424}{dB}$
    from below.
\end{theorem}
\begin{proof}
The proof can be found in Appendix~\ref{app:firstOrderBICM}.
\end{proof}

The results for BPSK, QPSK (2-PAM$\times$2-PAM), and 16-QAM (4-PAM$\times$4-PAM), as presented in Table~\ref{tab:ebnoMin-CM}, match with the Theorem.

It is somewhat surprising that the loss with respect to coded modulation at low $\snr$ is bounded.
The loss represents about $\units{1.25}{dB}$ with respect to the classical CM limit, namely $\ebnoZ = \units{-1.59}{dB}$. In the next section, we examine in detail the precise extent to which this loss translates into an equivalent loss in power. We will do so by allowing for simultaneous variations in power and bandwidth and conclude that using BICM over a fixed modulation for a large range of signal-to-noise ratio values, where the transmission rate is adjusted by changing the code rate, needs not result in a large loss with respect to more optimal schemes, where both the rate and modulation change. Additionally, this loss can be traded off against a large bandwidth reduction.

\section{Bandwidth and Power Trade-off}
\label{sec:bandwidthPower}

In the previous section we computed the first coefficients of the Taylor expansion of the CM and BICM capacities around $\snr = 0$. In this section we use these coefficients to determine the trade-off between power and bandwidth in the low-power regime. We will see how part of the power loss incurred by BICM can be traded off against a large bandwidth reduction.

The data rate transmitted across a Gaussian channel is determined by two physical variables: the power $P$, or energy per unit time, and the bandwidth $W$, or number of channel uses per unit time. In this case, the signal-to-noise ratio $\snr$ is given by  $\snr = P/(N_0 W)$, where $N_0$ is the noise spectral density.
Then, the capacity measured in bits per unit time is the natural figure of merit for a communications system. With only a constraint on $\snr$, this capacity is given by $W\log(1+\snr)$. For low $\snr$, we have that
\begin{equation}
    W\log\biggl(1+\frac{P}{N_0W}\biggr) = \frac{P}{N_0} -\frac{P^2}{2 N_0^2W}+ \Ord\biggl(\frac{P^3}{N_0^3W^2}\biggr).
\end{equation}
Similarly, for coded modulation systems with capacity $\capCM$, we have
\begin{equation}\label{eq:39}
    {\capCM} W = \ca\frac{P}{N_0} + \cb\frac{P^2}{N_0^2W}+ \Ord\Biggl(\frac{P^{5/2}}{N_0^{5/2}W^{3/2}}\Biggr).
\end{equation}

Following Verd{\'u} \cite{verdu2002sew}, we consider the following scenario. Let two alternative transmission systems with respective powers $P_i$ and bandwidths $W_i$, $i = 1,2$, achieve respective capacities per channel use $\capc_i$. The corresponding first- and second-order expansion coefficients are denoted by $\cao, \cbo$ for the first system, and $\cat, \cbt$ for the second. A natural comparison is to fix a power ratio $\Delta P = P_2/P_1$ and then solve for the corresponding bandwidth ratio $\Delta W = W_2/W_1$ so that the data rate is the same, that is $\capc_1W_1 = \capc_2W_2$. For instance, option 1 can be QPSK modulation and option 2 use of a high-order modulation with BICM.

\subsection{An Approximation to the Trade-off}

When the capacities $\capc_1$ and $\capc_2$ can be evaluated, the exact trade-off curve $\Delta W(\Delta P)$ can be computed. For low power, a good approximation is obtained by keeping the first two terms in the Taylor series. Under this approximation, we have the following result.

\begin{theorem}\label{thm:powerBandwidthTradeOff}
In a neighbourhood of $\snr_1 = 0$ the capacities in bits per second, $\capc_1W_1$ and $\capc_2W_2$ are equal when the expansion factors $\Delta P$ and $\Delta W$ are related as
\begin{align}\label{eq:deltaWAWGN}
    \Delta W = \frac{\bigl(\cbt\snr_1 + \ord(\snr_1)\bigr) (\Delta P)^2}{\cao  + \cbo \snr_1 + \ord(\snr_1) - \cat \Delta P},
\end{align}
for $\Delta W$ as a function of $\Delta P$ and, if $\cat \neq 0$,
\begin{align}\label{eq:deltaPAWGN}
    \Delta P &= \frac{\cao}{\cat } + \biggl(\frac{\cbo}{\cat } -\frac{\cbt\cao^2}{\cat^3 \Delta W}\biggr)\snr_1 + \ord(\snr_1),
\end{align}
for $\Delta P$ as a function of $\Delta W$.
\end{theorem}
\begin{proof}
The proof can be found in Appendix~\ref{app:proofPowerBandwidthTradeOff}.
\end{proof}

Remark that we assume $\snr_1\to 0$. As a consequence, replacing the value of $\Delta P$ from Eq.~\eqref{eq:deltaPAWGN} into Eq.~\eqref{eq:deltaWAWGN} gives $\Delta W = \Delta W\bigl(1+\ord(\snr_1)\bigr)$,
which is not exact, but
valid within the approximation order.

The previous theorem leads to the following derived results. For simplicity, we drop the terms $\ord(\snr_1)$ and replace the equality signs by approximate equalities.

\begin{corollary}
For $\Delta P = 1$, we obtain
\begin{align}
    \Delta W \simeq \frac{\cbt\snr_1 }{\cao  + \cbo \snr_1 - \cat },
\end{align}
and for the specific case $\cao = \cat$,  $\Delta W \simeq \cbt/\cbo$.
\end{corollary}
The latter formula has also been obtained by Verd{\'u} \cite{verdu2002sew} as a ratio of wideband slopes.

As noticed in \cite{verdu2002sew}, the loss in bandwidth may be significant when $\Delta P = 1$. But this point is just one of a curve relating $\Delta P$ and $\Delta W$. For instance, with no bandwidth expansion we have
\begin{corollary}
For $\cao = \cat = 1$, and choosing $\Delta W = 1$, $\Delta P \simeq 1 + \bigl(\cbo -\cbt\bigr)\snr_1$.
\end{corollary}

For signal-to-noise ratios below \units{-10}{dB}, the approximation in Theorem~\ref{thm:powerBandwidthTradeOff} seems to be very accurate for ``reasonable'' power or bandwidth expansion ratios. A quantitative definition would lead to the problem of the extent to which the second order approximation to the capacity is correct, a question on which we do not dwell further.

Another example concerns the effect of fully-interleaved fading.
Let us consider a Nakagami-$\gf$ fading model, such that the squared fading coefficient $\chi_k = |h_k|^2$ follows a gamma distribution.
The parameter $\gf$ is a real positive number, $0<\gf<\infty$. 
Using the values of the moments of the gamma distribution, $\E[\chi] = 1$, and $\E[\chi^2] = 1+1/\gf$, we have that $\ca = \ca^\text{AWGN}$ and $\cb = \bigl(1+ \frac{1}{\gf}\bigr)\cb^\text{AWGN}$. Therefore
\begin{corollary}
Consider a modulation set $\Xcal$ with average unit energy and used with power $P$, bandwidth $W$, and signal-to-noise ratio $\snr$; its capacity in absence of fading is characterized at low $\snr$ by the coefficients $\ca = 1$ and $\cb$.
When used in the Nakagami-$\nu$ channel with power $P_\nu$ and bandwidth $W_\nu$, if $P_\nu = P$,
    $W_\nu = W\bigl(1+ \frac{1}{\gf}\bigr)$,
and if $W_\nu = W$,
    $P_\nu = P\bigl(1 - \frac{\cb}{\gf}\snr\bigr)$.
\end{corollary}
As expected, for unfaded AWGN, when $\gf \to \infty$, there is no loss.
%
Rayleigh fading ($\nu = 1$) incurs in a bandwidth expansion of a factor 2 if the power is to be fixed. On the other hand, if bandwidth is kept unchanged, there is a power penalty in dB of about $10\log_{10}(1 - \cb\snr) \simeq -10\cb\snr/\log 10 \simeq \units{-4.343\cb\snr}{dB}$, a negligible amount to all practical effects since $\snr \to 0$. The worst possible fading is $\nu \to 0$, which requires an unbounded bandwidth expansion or an unlimited power penalty. 

\subsection{Trade-off for BICM}

\begin{figure}[htb]
  \centering
  \includegraphics[width=0.9\columnwidth]{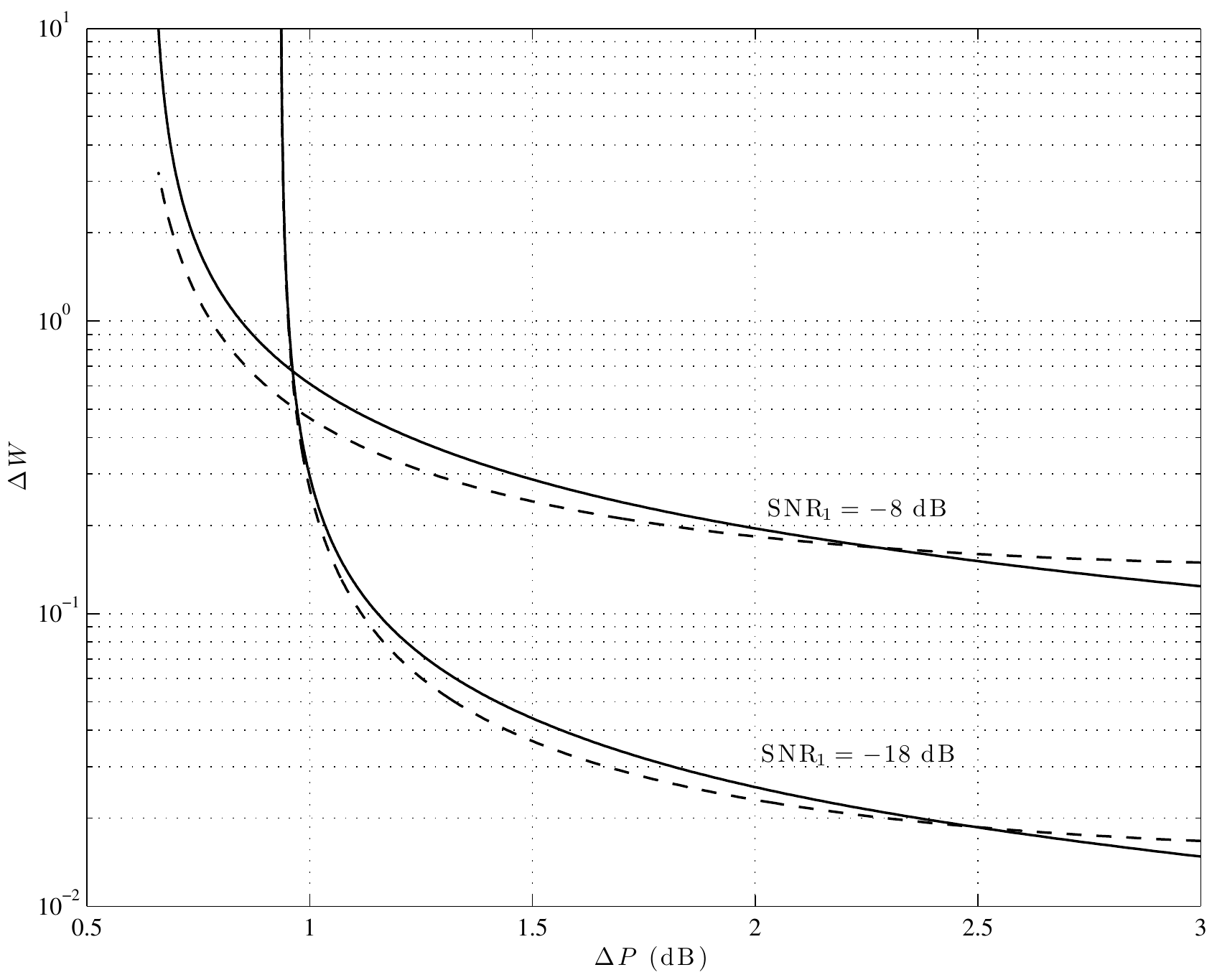}
  \caption{Trade-off between $\Delta P$ and $\Delta W$ between QPSK and 16-QAM with Gray mapping. Solid lines correspond to the exact tradeoff, while dashed lines correspond to the low-SNR tradeoff.}
  \label{fig:t20070129-16qam}
\end{figure}

The trade-off between power and bandwidth can also be applied to determine the expansion factors when BICM with a non-binary modulation is used rather than, say, QPSK modulation.
Fig.~\ref{fig:t20070129-16qam} shows the trade-off between power and bandwidth expansion factors when BICM over 16-QAM with Gray mapping is used, having taken QPSK as the reference transmission method. Results are presented for two values of the signal-to-noise ratio for the QPSK baseline. The exact result, obtained by using the exact formulas for $\capCM$ and $\capBICM$, respectively Eqs.~\eqref{eq:capCMAWGN} and~\eqref{eq:capBICM}, is plotted along the result by using Theorem~\ref{thm:powerBandwidthTradeOff}.

As expected, for very low values of $\snr$, the curve for $\Delta W$ diverges as $\Delta P$ approaches the value $\frac{\cao}{\cat} = \frac{1}{0.8}$, or \units{0.97}{dB}. This is in line with the fact that the minimum energy per bit required for 16-QAM/BICM is \units{-0.63}{dB}, as given in Table~\ref{tab:ebnoMin-CM}. Close to this limit, small improvements in power efficiency are extremely costly in bandwidth resources.
On the other hand, this loss may be accompanied by a significant reduction in bandwidth, which might be of interest in some applications. For instance, a loss of \units{2.4}{dB} from the baseline at \units{-18}{dB} requires a tiny fraction of the original bandwidth, about 2\%.

Concerning the last point, the results are exclusive to BICM and
the same analysis can be applied to a single transmission method with coefficients $\ca$ and $\cb$, trading off power against bandwidth. In this case, for a given $\Delta P$ we would have
\begin{align}
    \Delta W \simeq \frac{\cb\snr_1 (\Delta P)^2}{\ca(1-\Delta P)  + \cb \snr_1}. 
\end{align}
Using QPSK ($\ca = 1$, $\cb = -\frac{1}{2}$) and for $\snr = \units{-18}{dB}$ a loss of \units{2.4}{dB} is linked to using only 3\% of the original bandwidth. We see that QPSK is slightly more inefficient than BICM/16-QAM in using the bandwidth, the reason being that it has a lower coefficient $\cb$, $-0.5$ instead of $-0.16$. To any extent, it should not be surprising that communication in the wideband regime can be inefficient in using the bandwidth, since we are working in a regime where the main limitation is in power.


For signal-to-noise ratios larger than those reported in the figure, the assumption of low $\snr$ loses its validity and the results derived from the Taylor expansion are no longer accurate.

\section{Conclusions}
\label{sec:conclusions}

In this paper, we have computed the first two derivatives of the constrained capacity at zero $\snr$ for rather general modulation sets, and used the result to characterize analytically the bahaviour of BICM in the low-power regime. For binary reflected Gray mapping, the capacity loss at low $\snr$ with respect to coded modulation is shown to be bounded by approximately $\units{1.25}{dB}$.
This fact may be useful for the design of systems operating at low signal-to-noise ratios.

Moreover, we have determined the trade-off at low $\snr$ between power penalty and bandwidth expansion between two alternative systems. The trade-off presented here generalizes Verd{\'u}'s analysis of the wideband regime, where the bandwidth expansion for a fixed power was estimated. We have shown that no bandwidth expansion may be achieved at a negligible (but non-zero) cost in power. A similar trade-off between power penalty and bandwidth expansion for general Nakagami-$\nu$ fading has been computed, with similar conclusions as in the point above: bandwidth expansion may be large at no power cost, but absent at a tiny power penalty. We have applied the trade-off to a comparison between QPSK and 16-QAM.

\newpage
\appendices

\section{Proof of Proposition \ref{prop:proofCBICM}}
\label{app:proofCBICM}

By definition, the BICM capacity is the sum over $i = 1,\dotsc,m$ of the mutual informations $I(B_i;\Zz)$. We rewrite this mutual information as
\begin{align}
    I(B_i;\Zz) &= \frac{1}{2}\sum_{b\in\{0,1\}}\E\Biggl[\log \frac{\sum_{\sss'\in\Xcali} \pzxp}{\frac{1}{2}{\sum_{\sss'\in\Xcal} \pzxp}}\Biggr]\\
    &= \frac{1}{2}\sum_{b\in\{0,1\}}\E\Biggl[\log \Biggl(\frac{\sum_{\sss'\in\Xcali}\frac{2}{|\Xcal|} \pzxp}{\pzx}\frac{\pzx}{\frac{1}{2}{\sum_{\sss'\in\Xcal} \frac{2}{|\Xcal|}\pzxp}}\Biggr)\Biggr],
\end{align}
where we have modified the variable in the logarithm by including a factor $\frac{2}{|\Xcal|}\pzx$ in both numerator and denominator. Splitting the logarithm,
\begin{align}
    I(B_i;\Zz) = \frac{1}{2}&\sum_{b\in\{0,1\}}\E\Biggl[\log \frac{\sum_{\sss'\in\Xcali} \frac{2}{|\Xcal|}\pzxp}{\pzx}\Biggr]\notag\\
& + \frac{1}{2}\sum_{b\in\{0,1\}}\E\Biggl[\log \frac{\pzx}{\frac{1}{|\Xcal|}{\sum_{\sss'\in\Xcal} \pzxp}}\Biggr].
\end{align}

For fixed $b$, the quantity
\begin{equation}
    -\E\Biggl[\log \frac{\sum_{\sss'\in\Xcali} \frac{2}{|\Xcal|}\pzxp}{\pzx}\Biggr]
\end{equation}
is
the mutual information achievable by using equiprobable signalling in the set $\Xcali$, $\capCMi^\text{u}$,
and, similarly, the quantity
\begin{equation}
    \E\Biggl[\log \frac{\pzx}{\frac{1}{|\Xcal|}{\sum_{\sss'\in\Xcal} \pzxp}}\Biggr]
\end{equation}
is the mutual information achieved by equiprobable signalling in $\Xcal$, $\capCM^\text{u}$.

\newpage
\section{Linear Expansion Capacity}
\label{app:proofAsymptotic}

We start with \eqref{eq:c_snr} and use Lagrange's inversion formula. The inversion formula transforms a function
\begin{equation}
    \capc = f_1(\snr)
\end{equation}
into its inverse
\begin{equation}
    \snr = f_2(\capc).
\end{equation}
We do an expansion around $\snr = 0$, which is also $\capc = 0$. Applied to our case, the inversion formula becomes
\begin{align}
    \snr = \frac{\snr}{f_1(\snr)} \biggl|_{\snr\to 0}\capc ~+~  \frac{1}{2}\frac{d}{d\snr}\biggl(\frac{\snr}{f_1(\snr)}\biggr)^2 \biggl|_{\snr\to 0} \capc^2 + \Ord(\capc^3).
\end{align}
Using the expansion in \eqref{eq:c_snr}, after some simplifications we get
\begin{align}
    \snr
    &= \frac{\log2}{\ca}\capc  - \frac{\cb\log^2 2}{\ca^3}\capc^2 + \Ord(\capc^3).
\end{align}
Letting $\snr=\capc\,\ebno$ and rearranging we obtain
\begin{align}
    \ebno  &= \frac{\log2}{\ca}  - \frac{\cb\log^2 2}{\ca^3}\capc + \Ord(\capc^2),
\end{align}
which leads to
\begin{align}
    \capc  &=   - \frac{\log^2 2\cb}{\ca^3}\left(\ebno - \frac{\log2}{\ca}\right)+ \Ord\left(\left(\ebno-\frac{\log2}{\ca}\right)^2\right), \nonumber
\end{align}
and hence the desired result.

\newpage
\section{CM Capacity Expansion at Low SNR}
\label{app:cmGaussian-prelov}
The assumption that the constellation moments are finite implies that
    $\E\bigl[|\Ss|^{2+\alpha}\bigr] < \infty$
for $\alpha > 0$. Therefore, as $\snr\to 0$, for $\mu> 0$ the technical condition
\begin{equation}
    \snr^{2+\alpha}\E\bigl[|\Ss|^{2+\alpha}\bigr] \leq (-\log \sqrt{\snr})^\mu,
\end{equation}
necessary to apply Theorem~5 of \cite{prelov04:soa} holds.

\newcommand{\sr}{\sss^{(r)}}
Let us define a $2\times 1$ vector $\sr = (\sss_r \,\,\sss_i)^T$, with components the real and imaginary parts of symbol $\sss$, respectively denoted by $\sss_r$ and $\sss_i$. The covariance matrix of $\sr$, denoted by $\cov(\Ss)$, is given by
\begin{align}
    \cov(\Ss) 
    &= \begin{pmatrix} \E[(\Ss_r - \sAvr)^2] & \E\bigl[(\Ss_r - \sAvr)(\Ss_i - \sAvi)\bigr]  \\ \E\bigl[(\Ss_r - \sAvr)(\Ss_i - \sAvi)\bigr] &\E[(\Ss_i - \sAvi)^2] \end{pmatrix},
\end{align}
where $\sAvr$ and $\sAvi$ are the mean values of the real and imaginary parts of the constellation.

Theorem~5 of \cite{prelov04:soa} gives $\ca = \Tr(\cov(\Ss))$ and $\cb = -\Tr(\cov^2(\Ss))$, or
\begin{align}
    \ca &= \E[(\Ss_r - \sAvr)^2] + \E[(\Ss_i - \sAvi)^2] \label{eq:caproof}\\
    \cb &= -\Bigl(\E^2[(\Ss_r - \sAvr)^2] + \E^2[(\Ss_i - \sAvi)^2] + 2\E^2\bigl[(\Ss_r - \sAvr)(\Ss_i - \sAvi)\bigr] \Bigr)\label{eq:cbproof}.
\end{align}
The coefficient $\ca$ coincides with that in Eq.~\eqref{eq:c1c2CM-1}.

As for $\cb$, let us add a subtract a term $\E[(\Ss_r-\sAvr)^2]\E[(\Ss_i-\sAvi)^2]$ to Eq.~\eqref{eq:cbproof}. Then,
\begin{align}
    \cb = -\Bigl(\tfrac{1}{2}\E^2&[(\Ss_r - \sAvr)^2] + \tfrac{1}{2}\E^2[(\Ss_i - \sAvi)^2] +\E[(\Ss_r-\sAvr)^2]\E[(\Ss_i-\sAvi)^2]\notag \\ &+\tfrac{1}{2}\E^2[(\Ss_r - \sAvr)^2] + \tfrac{1}{2}\E^2[(\Ss_i - \sAvi)^2]\notag\\&- \E[(\Ss_r-\sAvr)^2]\E[(\Ss_i-\sAvi)^2]+ 2\E^2\bigl[(\Ss_r - \sAvr)(\Ss_i - \sAvi)\bigr] \Bigr),
\end{align}
which in turn can be written as
\begin{align}
    \cb = -\frac{1}{2}\Bigl(\E^2\bigr[|\Ss-\sAv|^2\bigr]+\bigl|\E[(\Ss-\sAv)^2]\bigr|^2\Bigr),
\end{align}
a form which coincides with Eq.~\eqref{eq:c1c2CM-2}, by noting that
\begin{align}
    \E\bigr[|\Ss-\sAv|^2\bigr] &= \E\bigr[|\Ss|^2\bigr]-|\sAv|^2 = \muTwoConst - \bigl|\meanConst\bigr|^2 \\
    \E[(\Ss-\sAv)^2] &= \E[\Ss^2]-\sAv^2 = \muTwoBisConst - \meanConstSqrd .
\end{align}

\newpage
\section{Proof of Theorem \ref{prop:c1c2BICM}}
\label{app:c1c2BICM}

In Eq.~\eqref{eq:capBICM} for the BICM capacity, the summands $\capCM$ and $\capCMi$ admit each a Taylor expansion given in Theorem~\ref{prop:c1c2CM}. Hence,
\begin{align}
    \ca &= \sum_{i=1}^m \frac{1}{2}\sum_{b=0,1} \Bigl(1 - \bigl(\mu_2(\Xcali)-|\sbi|^2\bigr)\Bigr) \\
    &= \sum_{i=1}^m\Biggl( \biggl(1 - \frac{1}{2}\sum_{b=0,1} \mu_2(\Xcali)\biggr) + \frac{1}{2}\sum_{b=0,1} |\sbi|^2\Biggr)\\
    &= \sum_{i=1}^m\Biggl( \biggl(\sum_{s\in\Xcal}\frac{1}{|\Xcal|}|s|^2 - \sum_{s\in\Xcal}\frac{1}{|\Xcal|}|s|^2\biggr) + \frac{1}{2}\sum_{b=0,1} \sum_{s\in\Xcali}\frac{2}{|\Xcal|}|\sbi|^2\Biggr)\\
    &= \sum_{i=1}^m\frac{1}{2}\sum_{b=0,1} |\sbi|^2,
\end{align}
since $\frac{1}{2}\sum_{b=0,1} \mu_2(\Xcali) = \mu_2(\Xcal) = 1$ by construction.

As for $\cb$, it follows from a similar application of Theorem~\ref{prop:c1c2CM}.


\newpage
\section{First-order Coefficient for BICM with Gray mapping}
\label{app:firstOrderBICM}

For $M$-PAM, the Gray mapping construction makes $\sbi= 0$, for $b=0,1$ and all bit positions except one, which we take with no loss of generality to be $i = 1$. Therefore,
\begin{equation}
    \ca = \frac{1}{2}\bigl|\mu_1(\Xcal_1^0)\bigr|^2+\frac{1}{2}\bigl|\mu_1(\Xcal_1^1)\bigr|^2 = \bigl|\mu_1(\Xcal_1^0)\bigr|^2 = \bigl|\mu_1(\Xcal_1^1)\bigr|^2.
\end{equation}
The last equalities follow from the symmetry between 0 and 1.

Symbols lie on a line in the complex plane with values
    $\pm\beta\bigl(1,3,5,\dotsc,M-1)$, with $\beta$ a normalization factor $\beta^2 = 3/(M^2-1)$.
This factor follows by setting $2n = M$ in the formula
    $\frac{1}{n}\sum_{i=1}^n (2i-1)^2 = \frac{1}{3}((2n)^2-1)$,
The average symbol has modulus
    $|\mu_1(\Xcal_1^0)| = \beta \frac{M}{2}$,
and therefore
\begin{equation}
    \ca = \bigl|\mu_1(\Xcal_1^0)\bigr|^2 = \frac{3\cdot M^2}{4(M^2-1)}.
\end{equation}

Extension to $M^2$-QAM is clear, by taking the Cartesian product along real and imaginary parts. Now, two indices $i$ contribute, each with an identical form to that of PAM. As the energy along each axis of half that of PAM, the normalization factor $\beta_\text{QAM}^2$ also halves and overall $\ca$ does not change.

\newpage
\section{Determination of the Power and Bandwidth Trade-Off}
\label{app:proofPowerBandwidthTradeOff}

In order to have the same capacities bandwidth and/or power must change to account for the difference in capacity, so that
\begin{align}
    \cao \frac{P_1}{N_0} + \cbo \frac{P_1^2}{W_1 N_0^2} + \ord(W_1\snr_1^2)=
    \cat \frac{P_2}{N_0} + \cbt \frac{P_2^2}{W_2 N_0^2} + \ord(W_2\snr_2^2).
\end{align}

Simplifying common factors, we obtain
\begin{align}
    &\cao  + \cbo \snr_1 + \ord(\snr_1)= \cat \frac{P_2}{P_1} + \bigl(\cbt + \ord(\snr_1)\bigr)\frac{P_2^2}{P_1^2}\frac{W_1}{W_2}\snr_1.
\end{align}
Or, with the definitions $\Delta P = P_2/P_1$, and $\Delta W = W_2/W_1$,
\begin{align}
    \cao  + \cbo \snr_1 + \ord(\snr_1) = \cat \Delta P  + \bigl(\cbt\snr_1 + \ord(\snr_1)\bigr)\frac{(\Delta P)^2}{\Delta W},
\end{align}
and
\begin{align}\label{eq:app54}
    \Delta W &= \frac{\bigl(\cbt\snr_1 + \ord(\snr_1)\bigr) (\Delta P)^2}{\cao  + \cbo \snr_1 + \ord(\snr_1) - \cat \Delta P}.
\end{align}
This equation gives the trade-off between $\Delta P$ and $\Delta W$, for a fixed (small) $\snr_1$, so that the capacities of scenarios 1 and 2 coincide.

Next we solve for the inverse, i.\ e.\ for $\Delta P$ as a function of $\Delta P$. First, let us define the quantities $a = \cbt\snr_1 + \ord(\snr_1)$ and $b = \cao  + \cbo \snr_1 + \ord(\snr_1)$. Then, rearranging Eq.~\eqref{eq:app54} we have $a(\Delta P)^2 + \cat \Delta W\Delta P - b\Delta W = 0$ and therefore
\begin{align}
    \Delta P &= \frac{-\cat \Delta W \pm \sqrt{(\cat \Delta W)^2 + 4ab\Delta W }}{2a} \\
    &= \frac{\cat \Delta W}{2a}\biggl(-1 \pm \sqrt{1 + \frac{4ab}{\cat^2 \Delta W} }\biggr)
    \label{eq:29}.
\end{align}
Often we have $\cbt < 0$, and then the negative root is a spurious solution. We choose then the positive root. Since $ab$ is of order $\snr_1$, we can use the Taylor expansion $(1+4t)^{1/2} = 1 + 2t - 2t^2+\ord(t^2)$, to write
\begin{align}
    \Delta P &= \frac{\cat \Delta W}{2a}\biggl(\frac{2ab}{\cat^2 \Delta W} - \frac{2a^2b^2}{\cat^4(\Delta W)^2}\biggr) \\
    &= \frac{b}{\cat} - \frac{ab^2}{\cat^3\Delta W}.
\end{align}
Since $\snr_1\to 0$, we group the non-linear terms in $\snr_1$ and so get
\begin{align}
    \Delta P &= \frac{\cao  + \cbo \snr_1 }{\cat} - \frac{\cbt\cao^2\snr_1 }{\cat^3\Delta W}+ \ord(\snr_1) \\
    &= \frac{\cao}{\cat } + \biggl(\frac{\cbo}{\cat } -\frac{\cbt\cao^2}{\cat^3 \Delta W}\biggr)\snr_1 + \ord(\snr_1) .
\end{align}

\newpage


\end{document}